\documentclass[aps,prl,twocolumn,showpacs,groupedaddress]{revtex4}
\usepackage{graphicx}

\begin{document}

\title{Discrepancies in Determinations of the Ginzburg-Landau Parameter}

\author{TA Girard} \affiliation{Centro de
F\'isica Nuclear, Universidade de Lisboa, 1649-003 Lisbon, Portugal}
\author{S. Figueiredo}
\affiliation{Centro de F\'isica Nuclear, Universidade de Lisboa,
1649-003 Lisbon, Portugal}

\date{\today}

\begin{abstract}
Long-standing discrepancies within determinations of the
Ginzburg-Landau parameter $\kappa$ from supercritical field
measurements on superconducting microspheres are reexamined. The
discrepancy in tin is shown to result from differing methods of
analyses, whereas the discrepancy in indium is a consequence of
significantly differing experimental results. The reanalyses however
confirms the lower $\kappa$ determinations to within experimental
uncertainties.
\end{abstract}

\pacs{$74.25.-q, 74.25.Nf, 74.62.-c$}

\maketitle

\section{I. INTRODUCTION}

The Ginzburg-Landau parameter of a superconductor, $\kappa$, is
generally defined as the ratio of its magnetic penetration depth
($\lambda$) to its coherence length ($\xi$) at the thermodynamical
critical temperature ($T_{c}$). The parameter relates the two
fundamental length scales of the material's superconducting phase,
distinguishes whether the phase transition is first- or
second-order, and characterizes the material's response to applied
magnetic fields in the superconducting state. Theoretically, for
clean materials, it is given by the BCS $\kappa$ = 0.96
$\lambda_{L}$(0)/$\xi_{0}$ where $\lambda_{L}$(0) is the London
penetration depth at zero temperature and $\xi_{0}$ is the Pippard
coherence length \cite{tinkh,degen}.

The parameter, although often referred to, is infrequently measured
and little tabulated owing to its dependence on sample purity and
structure. Where measured,  $\kappa$ has been determined
\cite{degen} from independent measurements of $\lambda$ and the
thermodynamic critical field $H_{c}$, as well as from magnetization
measurements on thin films and foils of various materials
\cite{chang,miller}. For type-I materials ($\kappa <$ 1/$\sqrt{2}$),
another technique has been from measurements of the supercritical
fields of microspheres. This technique has been used in determining
the $\kappa$ of aluminum \cite{sbc,cruz}, cadmium \cite{cruz2},
gallium \cite{fedkiroth}, mercury \cite{sbc,sc,burger}, indium
\cite{sbc,feder,fedkiroth2,sc2}, lead \cite{sbc,sc}, tin
\cite{sbc,feder,larrea,sc3,sc4}, thallium \cite{sbc} and zinc
\cite{cruz}, as well as a series of alloys \cite{sbc}. Inexplicably,
the results are generally $\sim$ factor 2 smaller than the thin
film/foil determinations as well as the BCS-defined $\kappa$, as
shown in Table I. More recent measurements, using a fast pulse
induction technique with thin foils, provide results in better
agreement with those from the microspheres \cite{valko}.

\begin{table}[h]
  \caption{comparison of Ginzburg-Landau parameter determinations;
  "-" denotes no experimental measurement.}\label{Table 1}
  \begin{tabular}{|c|c|c|c|c|c|c|}
\hline  & lead & thallium &  tin & indium & cadmium & aluminum \\
\hline $\kappa_{grains}$ & 0.25 & 0.076 & 0.086 & 0.066 & 0.012 & 0.013 \\
\hline $\kappa_{film/foil}$ & 0.34 & - & 0.15 & 0.13 & - & - \\
\hline $\kappa_{BCS}$ & 0.43 & 0.12 & 0.16 & 0.12 & 0.003 & 0.014\\
\hline
\end{tabular}
\end{table}

The discrepancy is further complicated by severe disagreements
between the microsphere reports themselves. This is shown in Fig. 1
for tin and indium ($\kappa \sim$ 0.1), two of the materials most
studied with this technique, where the closed (open) symbols of each
figure refer to the effective $\kappa_{sh}$($\kappa_{sc}$) from the
Refs. indicated, defined by

\begin{equation}\label{1}
\left\{\begin{array}{l}
     \sqrt{2} \kappa_{sh}(t) =  h_{sh}^{-2 }(t)\\
     \sqrt{2} \kappa_{sc}(t) =  h_{sc}(t)   , \\
\end{array}
\right.
\end{equation}

\noindent where $h_{sh}$ = $H_{sh}$/$H_{c}$ is the reduced
superheating field, $h_{sc}$ = $\eta h_{c2}$ is the reduced
supercooling field, $\eta(t=1)$ = 1.695, and $t$ = $T$/$T_{c}$ is
the reduced temperature. Although there is no complete theoretical
description of the temperature behavior of the supercritical fields
which spans the entire $T-H$ phase space, the parameters in
principle converge to $\kappa \equiv \kappa(1)$ provided the
microspheres are sufficiently large to prevent size effects which
arise when $\xi(t)$ becomes comparable to the dimensions of the
sample.

Whereas the respective $\kappa_{sc}(t)$ are seen in generally good
agreement, the $\kappa_{sh}(t)$ differ significantly, amounting to
as much as 30 G in either material. The $\kappa_{sh}(t)$ manifests
the parabolic behavior of h$_{c}(t)$; the near-linear behavior of
the $\kappa_{sc}(t)$, at least above t $\sim$ 0.4, is consistent
with magnetization measurements \cite{chang,auer}, with the
different curvatures reflecting the difference in dynamics between
the flux penetration and expulsion processes.

\begin{figure*}
\includegraphics[width=8 cm]{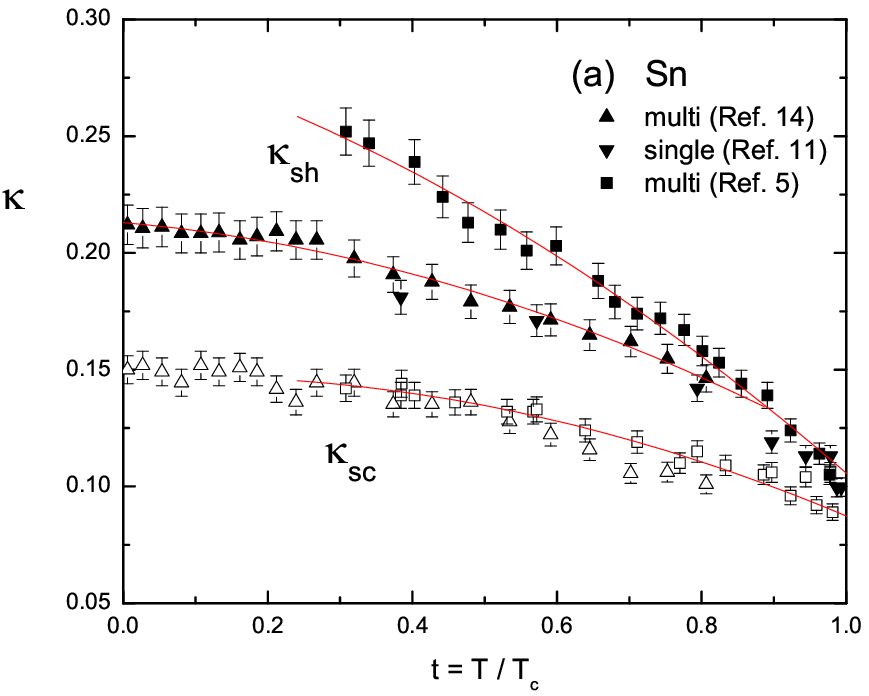}
\includegraphics[width=8 cm]{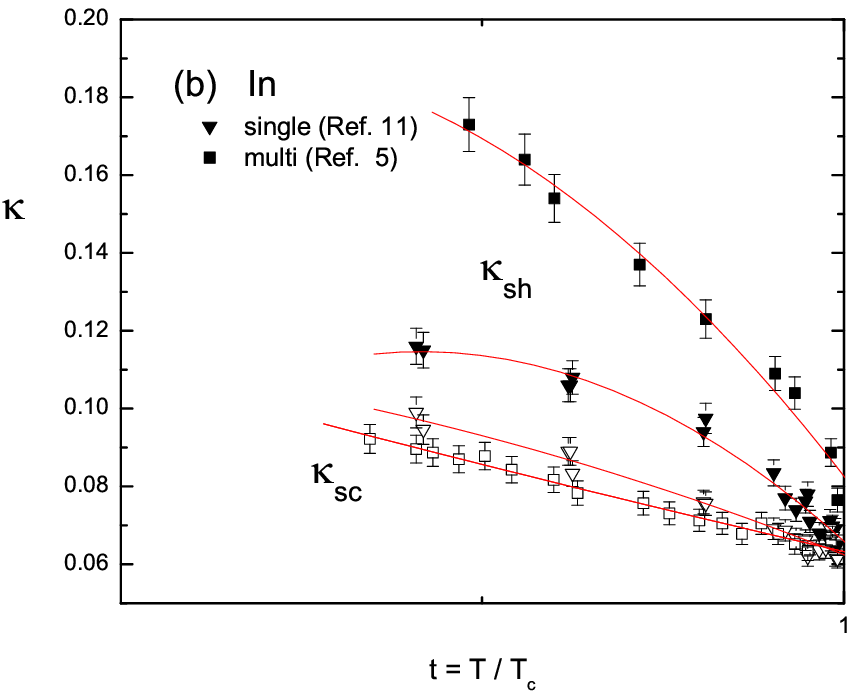}\\
\caption{(a) Compilation of $\kappa_{sh}(t)$, $\kappa_{sc}(t)$
extracted from measurements of the supercritical fields in tin
microspheres; (b) same as (a) for indium: closed symbols represent
superheating; open symbols, supercooling; lines represent polynomial
fits to the respective data sets.}
\end{figure*}

Eq. (1) implies the possibility that the higher  $\kappa_{sh}(t)$ of
Ref. \cite{sbc} were obtained with microspheres which failed to
achieve a full metastability, ie. which manifested lower
superheating fields. This could result from the presence of material
defects which effectively lower the transition field. The apparent
convergence of the different  $\kappa_{sh}(t)$, and the
$\kappa_{sh}(t)$, $\kappa_{sc}(t)$, at $t$ = 1 despite significant
differences at $t \ll$ 1, might then be the result of the
temperature dependences of both $\lambda$ and $\xi$, which diverge
as $t \rightarrow$ 1. The higher $\kappa_{sh}(t)$ of Ref. \cite{sbc}
might also arise from diamagnetic interactions between the
microspheres employed in the suspensions, absent in the single
sphere measurements of Ref. \cite{feder}, which would raise the
local fields so that the transitions appear at the lower (applied)
fields. Curiously however, the more recent suspension results
\cite{larrea} are in agreement with those of the single sphere.

Nevertheless, it would appear that the discrepancy has been
attributed to such extrinsic effects, and thusly disregarded: the
determinations of $\kappa$ have customarily proceeded by ignoring
the $\kappa_{sh}(t)$ to extract $\kappa$  from the $\kappa_{sc}(t)$
as $t \rightarrow$ 1 on the basis of Eq. (1) alone, even though
defects and diamagnetic interactions would also have impact on the
$h_{sc}$ measurements.

Although in evidence some thirty years ago, these discrepancies have
never been addressed in the literature insofar as we are aware. They
cast doubt on the determinations of $\kappa$ in Ref. \cite{sbc},
hence also on the results of Refs. \cite{cruz,cruz2,sc,sc2,sc3,sc4}
many of which are the only measurements existing for the given
material. They furthermore prevent the straightforward use of such
measurements in elaborating the behavior of $h_{sh}$ for $t \ll$ 1,
the precise temperature-dependence of which remains a question of
some theoretical interest \cite{lan}.

We here reexamine the analyses of the data, and clarify at least a
part of the problem. The data base and its reanalyses are described
in Sec. 2, with the results presented in Sec. 3. These are discussed
in Sec. 4, and conclusions presented in Sec. 5. Generally, we find
the discrepancy in tin to result from differing methods of analyses,
whereas the discrepancy in indium is a consequence of significantly
differing experimental results. Combination with the results from
$\kappa_{sc}(t)$ in each case yields results for $\kappa$ differing
only slightly from their previously-reported values, but with
somewhat larger uncertainties. Although we find no simple
explanation for the almost factor two difference between $\kappa$
determinations of this technique and those of the thin films/foils,
we observe that the latter are also obtained from $\kappa_{c2}$ =
$2^{-1/2}h_{c2}$ and that the difference with the microsphere
results is nearer a factor $\sim$ 1.7, suggesting a possible
mis-identification of $H_{c3}$ as $H_{c2}$ in the previous analyses.

\section{II.  DATA BASE}

The results by Feder and McLachlan (FM) for tin are obtained from
supercritical field measurements on single spheres of diameters 8,
21 and 48 $\mu$m, over the temperature range 0.4 $< t <$ 1
\cite{feder}. Both the 8 and 48 $\mu$m spheres exhibit strong size
and defect effects, and were not considered. The FM report on indium
is from measurements on single spheres of 8, 16 and 35 $\mu$m
diameters, and a powder of 10-50 $\mu$m spheres in volume
concentration of 17\%; we omit all but the 35 $\mu$m measurements as
a result of observed size and defect effects. The more recent tin
results of Larrea et. al. \cite{larrea}, from suspensions of
microspheres of diameter 33-40 $\mu$m at a volume filling factor of
25\%, are in general agreement with those of FM over 0.005 $< t <$
0.81. The errors in Fig. 1 represent 4\% uncertainties, and are for
reader convenience only: Larrea et. al. report errors of roughly
this level, and the reports of SBC and FM are assumed comparable
since no estimate is provided. There is also a report by Feder,
Kiser and Rothwarf \cite{fedkiroth2} with suspensions of 1-5 $\mu$m
tin spheres, which we however neglect since the estimated diameter
above which size effects are unimportant is $\sim$ 7.5 $\xi$(T), or
$\sim$ 5 $\mu$m at 0.95 $T_{c}$ \cite{feder}.

Smith, Baratoff and Cardona (SBC) reported results for tin from
measurements on suspensions of 5-15 $\mu$m diameter spheres
suspended to a volume filling factor of  $<$ 5\% over 0.3 $< t <$ 1,
and for indium from measurements on 1-10 $\mu$m microspheres with a
filling factor of $<$ 5\% over the same temperature range
\cite{sbc}.

\begin{figure*}
\includegraphics[width=8 cm]{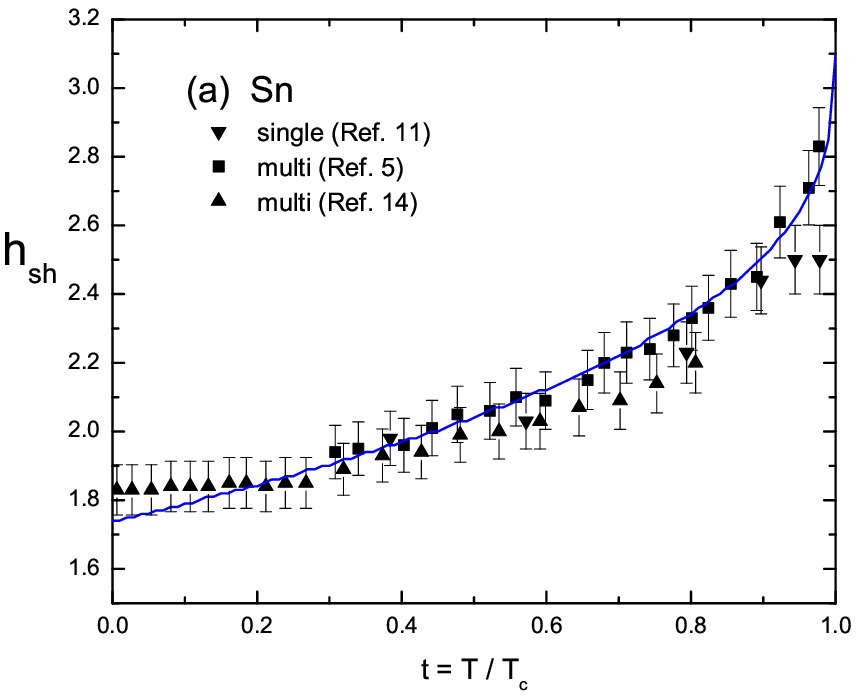}
\includegraphics[width=8 cm]{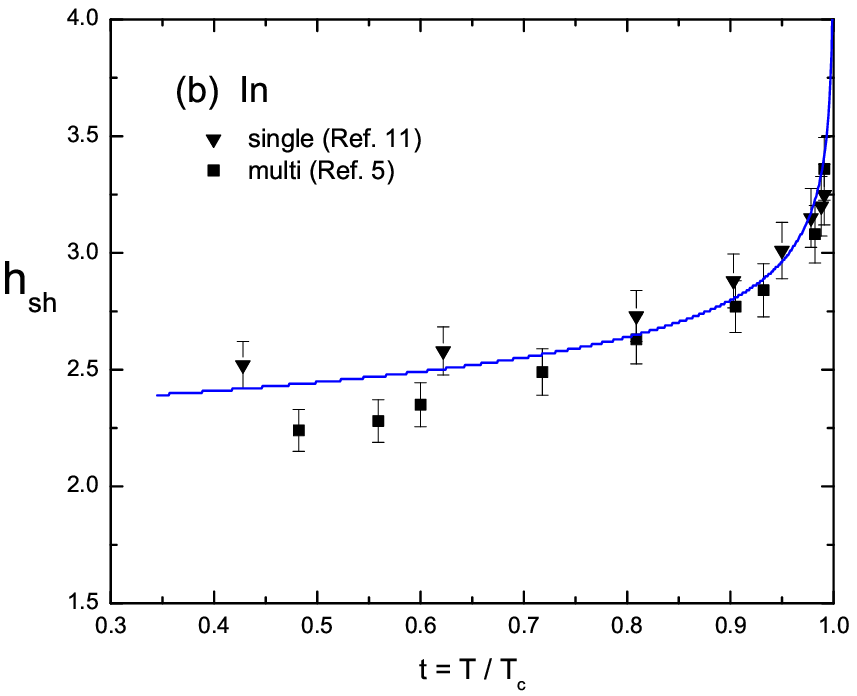}\\
\caption{\label{}((a) comparison of the regenerated SBC
{$h_{sh}(t)$} in tin with those of FM and Larrea et. al.,
regenerated via Eq. (1); (b) same as (a) for indium; lines are
calculated from Eq. (3).}
\end{figure*}

None of the previous reportings provide the actual supercritical
field measurements. For spheres, the $H_{sc}$ of Eq. (1) measured is
actually $H_{c3}$, the critical field for surface nucleation of the
superconductive state, which is favored over the bulk nucleation
field $H_{c2}$ \cite{tinkh,degen}. Moreover, the local superheating
field at the sphere equator is a demagnetization-enhanced 3/2 larger
than the laboratory field \cite{tinkh,degen}. For FM, Larrea et.
al., and $\kappa_{sc}(t)$ of SBC, the supercritical fields were
regenerated from the reported $\kappa_{sh}(t)$, $\kappa_{sc}(t)$
using:

\begin{equation}\label{2}
\left\{\begin{array}{l}
     h_{sh}^{lab}(t) = \frac{2}{3}h^{local}_{sh}(t) = (0.561) \kappa_{sh}^{-1/2}(t) \\
     h_{sc}^{lab}(t) = \eta h_{c2}^{local} = (2.40) \kappa_{sc}(t)   , \\
\end{array}
\right.
\end{equation}

\noindent where the prefactor in $h_{sh}$ includes the sphere
demagnetization. The factor $\eta$($t$=1) = 1.695 is valid only for
$t \sim$ 1, and in principle increases with decreasing temperature.
A variational lower bound  $\eta$($t$=0) = 1.925, has been obtained
from a microscopic analyses of a pure superconductor assuming
specular surface reflection \cite{ebneth}, but has not been used in
the analyses since Refs. \cite{sbc,feder,larrea} used only
$\eta$(1).

While the  $\kappa_{sc}(t)$ were obtained from Eq. (2), the SBC
$\kappa_{sh}(t)$ were obtained from a numerical integration of the
Ginzburg-Landau equations following Ginzburg \cite{ginz} with
correction for the demagnetization of the sphere. To retrieve the
$h_{sh}(t)$ of SBC, we best fit the upper curve of SBC Fig. 1 over
the interest range with a 7th order polynomial, and applied the
inverse transformation to the full  $\kappa_{sh}(t)$ of SBC Fig. 8.
This reproduces the results of SBC Fig. 8 to within 3\% over the 0.8
$< t <$ 1 as indicated in Fig. 2.

\section{III.  RESULTS}

For tin, the $h_{sh}(t)$ of all groups, shown in Fig. 2(a), are in
reasonable agreement, although the Larrea et. al. results appear
flat below $t$ = 0.2 and there is a general tendency for the SBC
results to be higher in the midrange temperatures. The rapid drop in
$h_{sh}$ below $t$ = 1 is consistent with nonlocal electrodynamics,
the line indicating the predicted behavior in the extreme nonlocal
limit \cite{tombara,sbc},

\begin{equation}\label{3}
     h_{sh}(t) = \sigma \chi^{1/4} \kappa^{-1/3} (1-t)^{-1/12}  , \\
\end{equation}

\noindent where $\sigma$ = 1.35 (1.42) for diffusive (specular
reflective) phonon surface scattering, and $\chi \sim$ 1. Eq. (3) is
in principle valid for 1 $>>$ 1-$t >>$ $\kappa^{2}$, or about 20\%
of the available temperature range.

The discrepancy between the SBC and FM/Larrea et. al. results for
tin appears to arise solely from the different analyses techniques.
This seems not the case for indium, as seen in Fig. 2(b). In this
case, the SBC fields do not appear to agree with those of FM, and
are generally lower by $\sim$ 2 . The disagreement increases with
decreasing temperature: for $t \sim$ 0.5, it amounts to $\sim$ 55 G.
For the same $h_{sh}$, the two analyses lead to a difference in
$\kappa_{sh}$ of order 20\% for $h_{sh}$ = 2.0 as shown in Fig. 3,
in agreement with the 16\% stated in Refs. 5 and 11; the difference
increases with decreasing $h_{sh}$, i.e. as $t \rightarrow$ 1.

\begin{figure}
\includegraphics[width=8 cm]{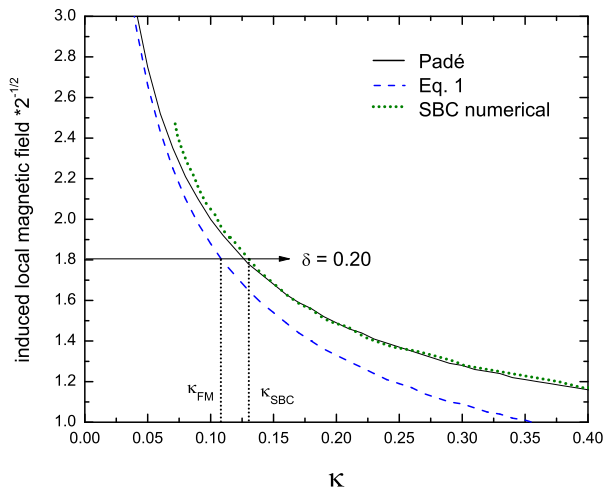}
\caption{\label{}the upper curve of SBC Fig. 1 together with Eq. (2)
and the Pad\'{e} $[2,2]$ approximant. Near the mid-range of reduced
superheating fields, the previous analyses techniques yield a
difference $\delta$ of $\sim$ 20\% in the determination of
$\kappa$.}
\end{figure}

The FM/Larrea et. al. analyses is the more customary, but that of
SBC would appear to be the more correct. FM have however argued
\cite{feder} that the numerical method employed by SBC, which allows
for only one-dimensional fluctuations of the order parameter, is too
large for $\kappa \sim$ 0.1; the SBC results therefore represent
only an upper limit. Permitting more than one degree of freedom in
the perturbation generally results in lower instability fields. On
the other hand, Eq. (1) represents only the first term in a general
expansion of $\kappa_{sh}$. Dolgert et. al. \cite{dolgert} have
reexamined the relation using the method of matched asymptotic
expansion, generating an expression for $\kappa_{sh}$ through 5th
order. The stability of the results was analyzed with respect to
both one- and two-dimensional perturbations, and the latter shown
not to lead to any additional destabilizing effects in the low-
limit. For present purposes, it is sufficient to reanalyze the
various $h_{sh}(t)$ with the Pad\'{e} [2,2] approximant
\cite{dolgert}, also shown in Fig. 3 and given by

\begin{equation}\label{4}
     H_{sh} = 2^{-3/4} \kappa^{-1/2}
     [\frac{1 + 5.444781 \kappa + 4.218101 \kappa^{2}}{1 + 4.781869 \kappa + 1.365523 \kappa^{2}}] . \\
\end{equation}

\noindent This agrees to within 1\% with more recent numerical
calculations for $\kappa <$ 1, which appear to differ only slightly
from the SBC curve above  $\kappa \sim$ 0.15. As evident in Fig. 3,
neither of the previous analyses appears in particularly good
agreement with Pad\'{e} in the region  $\kappa \sim$ 0.08.

Fig. 4(a) summarizes the combined results in tin using Eq. (4),
together with the respective  $\kappa_{sc}(t)$ determinations from
Eq. (2). Here, 4\% uncertainties are again shown except where
measurements overlap, in which case the standard mean and deviation
are given. The $\kappa_{sh}$ of Refs. \cite{feder} and \cite{larrea}
are raised, and coincide with those of Ref. \cite{sbc}. In
particular, both $\kappa_{sh}(t)$ and  $\kappa_{sc}(t)$ are seen not
to converge well at $t$ = 1.

The results of a least squares fitting are given in Table II,
together with the reported results of SBC and FM. The errors in this
work are fitting errors only, using $\kappa(t)$ =
m$_{1}$+m$_{2}$(1-$t$)$^{\rho}$ . As indicated, the
$\kappa_{sh}(t)$, $\kappa_{sc}(t)$ converge at $t$ = 1 to within
3$\sigma$, yielding $\kappa$(Sn) = 0.094 $\pm$ 0.006. Although the
difference with the previously-quoted results appears insignificant,
it occurs at the level of $\sim$ 4$\sigma$.

\begin{figure*}
\includegraphics[width=8 cm]{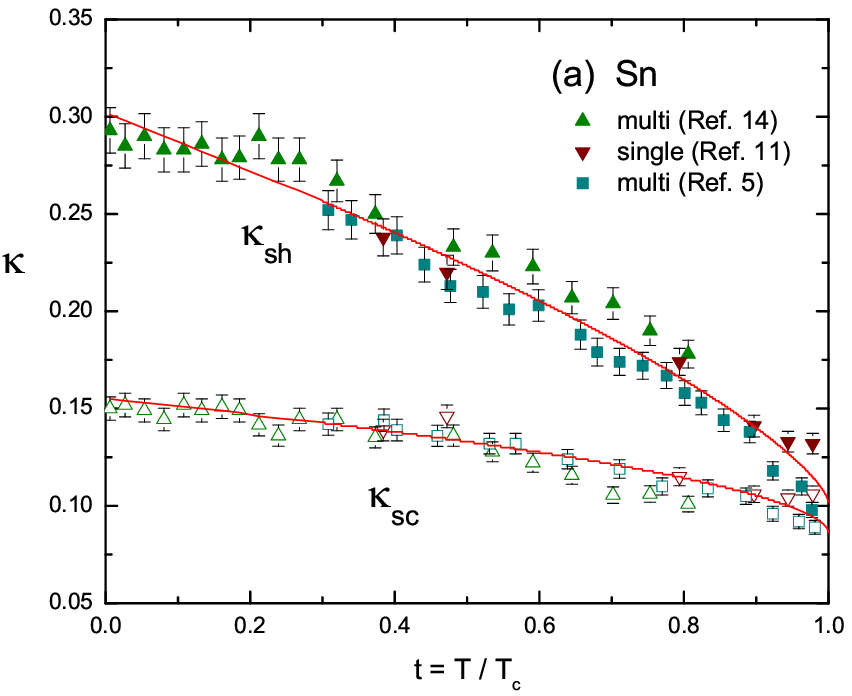}
\includegraphics[width=8 cm]{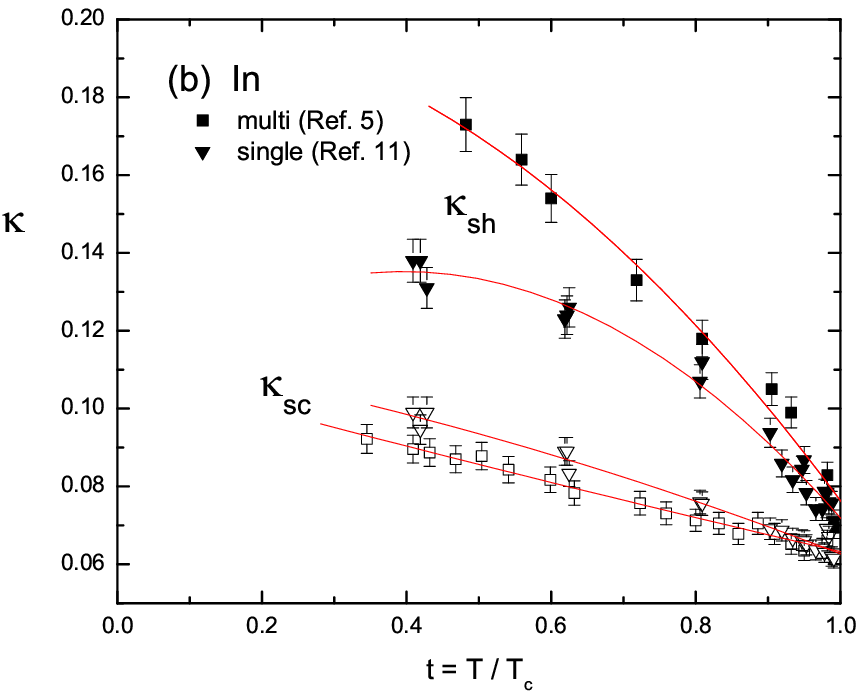}\\
\caption{\label{}((a) re-analyses of the combined $\kappa_{sh}$,
$\kappa_{sc}$ in tin via Eqs. (4) and (2). A least squares fit of
each yields the values for $\kappa$ shown in Table I; (b) same as
(a), for indium. The lower FM $\kappa_{sh}$ results for indium
suggest the grains of SBC not to have been fully metastable as a
result of defect presence; lines are polynomial fits to the
respective data sets.}
\end{figure*}

The similar reanalyses of $\kappa_{sh}(t)$ for indium using Eq. (4)
shows a continuing discrepancy below $t \sim$ 0.9. Although the SBC
$\kappa_{sh}(t)$ is little affected in the reanalyses, those of FM
are significantly increased above their Fig. 1 values. The implied
larger superheating fields of the FM determinations suggest that the
SBC suspension was in a mixed state and not fully superheated.
Nonetheless, both  $\kappa_{sh}(t)$ converge towards
$\kappa_{sc}(t)$ at $t$ = 1, as seen in Fig. 4(b), with $\kappa$(In)
= 0.064 $\pm$ 0.008 resulting from a (1-$t$)$^{\rho}$ fit. This is
shown in Table II in comparison with the results of SBC and FM. The
difference with previous determinations is well within error owing
to the fitting uncertainty. The analyses is however strongly
dependent on the data near $t$ = 1, with the larger uncertainty in
the indium result reflecting both the lack of data below $t \sim$
0.4 and measurement differences between the samples.

\begin{table}
  \caption{resume of $\kappa$ determinations in Sn and In from
  supercritical field measurements of small grains.}\label{Table 2}
  \begin{tabular}{|c|c|c|}
\hline  & tin & indium  \\
\hline SBC & 0.087 $\pm$ 0.002 & 0.060 $\pm$ 0.002  \\
\hline FM  & 0.093 $\pm$ 0.001 & 0.062 $\pm$ 0.001  \\
\hline this work (combined) & 0.094 $\pm$ 0.006 & 0.064 $\pm$ 0.008  \\
\hline
\end{tabular}
\end{table}

\section{IV.  Discussion}

The supercritical fields determinations of all reports are obtained
from hysteresis curves of the first order tin and indium
transitions. Different reports have however assumed different
definitions of the critical fields within the hysteresis curves, as
shown schematically in Fig. 5. As evident, these definitions alone
can lead to an appreciable variation in the reported {$\kappa(t)$}.
In general, the  $\kappa_{sh}$, $\kappa_{sc}$ parameters of FM
should be consistently higher than SBC, which is not observed in the
case of FM $\kappa_{sh}$.

In the case of single grain measurements, variations in the
supercritical fields arise mainly from the grain metallurgy. FM
performed careful investigations of defects as part of their study.
The hysteresis curves were measured as a function of the orientation
of the spheres in the applied field, and the  $\kappa_{sh}$,
$\kappa_{sc}$ taken from the direction yielding the lowest values
(i.e. highest $h_{sh}$, lowest $h_{sc}$). The results are strongly
dependent on $T_{c}$. In principle, for $T \sim T_{c}$, the
coherence length is sufficiently large that these effects are
negligible; at lower temperatures, the coherence length is smaller
than the defect and the effects are more noticeable. For this
reason, FM also employed $\kappa_{R}(t) \sim$
($h_{sc}$/$h_{sh}$)$^{2/3}$ as a comparison for {$\kappa_{sc}(t)$},
which has the advantage of being insensitive to $T_{c}$ since it is
independent of $H_{c}$.

In the case of the suspension measurements, the deformation of the
hysteresis from the rectangular single sphere cycle is due largely
to the interplay between defects and diamagnetic interactions
between the spheres \cite{hueber}. At lower temperatures, defects in
the sphere metallurgy may significantly reduce the superheating
capacity of the material, resulting in a lowered transition field
such as observed in the indium data of SBC. However, the defects
should also serve as nucleation centers for "early" normal (N)
$\rightarrow$ superconducting (S) transitions at lower temperatures,
which is not observed. On the other hand, the advantage of the
multisphere measurement lies in the averaging over the defect
contribution - the last sphere to transition to the normal state is
the most defect-free sphere of the suspension.

The recent work of Pe\~{n}aranda et. al. \cite{pen1} has
demonstrated significant diamagnetic interactions even for filling
factors as small as 5\%. These may vary from one suspension to
another of the same material as a result of inhomogeneous
distributions. Although the last sphere to transition in the S
$\rightarrow$ N case is effectively free of diamagnetic
interactions, this is not true for the N $\rightarrow$ S where
diamagnetic interactions are a maximum. As the field decrease
continues, the new N $\rightarrow$ S transitions enhance the local
diamagnetic contribution to the local fields of those spheres still
normal. This would suggest that the $\kappa_{sc}(t)$ determinations
with suspensions are lower than actual by some amount proportional
to the diamagnetic interactions, since the local field is higher. As
seen in Fig. 1(a), the SBC h$_{sc}(t)$ for tin is essentially the
same as those of the FM single sphere. For indium however, the SBC
$h_{sc}(t)$ are systematically lower than those of FM by $\sim$ 10\%
for $t \leq$ 0.8.

\begin{figure}
\includegraphics[width=8 cm]{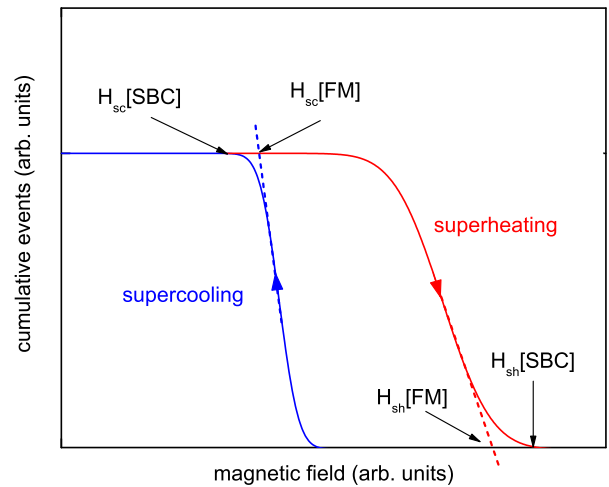}
\caption{\label{}hysteresis curve schematic, indicating the various
field definitions for $h_{sh}$, $h_{sc}$ employed by SBC and FM.}
\end{figure}

All results were moreover obtained by variation of the magnetic
field, rather than temperature, which as shown by Chaddah and Roy
\cite{chaddah} results in an enhanced $h_{sc}(t)$, apart from any
diamagnetic interaction, as a result of the thermal fluctuations
induced by the field changes themselves.

The variation of $\kappa$ with temperature suggests a variation of
the transition order with temperature, which has possibly important
ramifications since $\kappa$ is then less a fundamental property of
the superconductor than a simple ratio between the two
characteristic lengths in the description, both of which vary with
temperature and yield results consistent with the observed $\kappa$
determinations. Variation of the order of the transition with the
temperature is predicted in recent renormalization-based
reformulations of basic superconductive theory \cite{hove}, which
include fluctuations in the involved gauge and scalar fields, and
result in a dividing line between type-I and -II behavior at $\kappa
= 0.8/\sqrt{2}$ with a magnetic response which can be varied between
type-I and type-II simply by temperature change. This variation has
been seen in magnetization measurement of nitrogen-doped Ta
($\kappa$ = 0.665) \cite{auer}, but is otherwise unconfirmed.

\section{V.  Conclusions}

Reexamination of previous Ginzburg-Landau parameter determinations
from superheating field measurements of microspheres of tin and
indium shows the long-standing discrepancy in reported
$\kappa_{sh}(t)$ of tin to derive from the different analyses
methods employed, rather than whether single- or multi-grain
determinations, or grain metallurgy. Although the tin results of SBC
can be made to agree with those of FM (and Larrea et. al.), the
customary analyses via Eq. (2) is in disagreement with that
performed numerically. This is not the case of indium, where the
discrepancy reflects severe differences in the supercritical field
measurements themselves.

The similar reexamination of the supercooling field measurements
uncovers no divergences capable of explaining the discrepancies with
the measurements of Ref. \cite{chang,miller}. To the contrary, the
variation of $h_{sc}$ with $t$ appears to confirm the latter,
raising severe questions as to the true significance of $\kappa$  as
a descriptor of superconductors. The resolution of these is however
beyond the scope of this work.

The available $t <$ 1 data is of insufficient precision to permit
more than suggestions as to $\kappa$. Analyses of all superheating
results via the Pad\'{e} approximation yields a convergence of
$\kappa_{sh}(t)$, $\kappa_{sc}(t)$ to $\kappa$, with values for each
material only slightly different from those previously reported on
the basis of $h_{sc}(t)$ alone. These previous results however are
likely larger than actual as a result of diamagnetic interactions
between the superconducting sphere population. Apart from nonlocal
effects, the large  $\kappa_{sh}$ for $t <$ 0.8 is likely due to
defect presence, which causes nucleation of the normal state to
occur at fields below the maximal $H_{sh}$. These become unimportant
as $t \rightarrow$ 1, increasing $h_{sh}$ and generating a rapid
drop in $\kappa_{sh}$.

Whether the proper analyses is by Eq. (2a), numerical or Pad\'{e}
remains in question. Nevertheless, all of the materials analyzed in
Ref. \cite{sbc} (tin, indium, thallium, lead and mercury) follow
from the SBC Fig. 1; in contrast, the analyses of the remaining
Refs. has proceeded via Eq. (2), which is seen to underestimate the
respective $\kappa(t)$. In view of the significance of the
Ginzburg-Landau parameter, the uncertainties inherent to this
technique and the variation in experimentally-obtained results, and
technological advances over the last thirty years, careful
remeasurements of the supercritical fields over the full temperature
range, and their analysis within a definitive description of the
$\kappa-H_{sh}$ plane, would seem to be indicated.

The discrepancy in $\kappa$ between spheres and thin film/foil
determinations has been known for some decades, but to the best of
our knowledge remains unexplained. The thin film/foil results are in
fact a factor $\sim$ 1.7 larger than those from the microspheres;
they are also in better agreement with $\kappa$ derived from
$h_{c3}$ via $\kappa (t) = (1.695\surd 2)^{-1}h_{c3}(t)$, and also
agree in general with the lower temperature results of the
microspheres.

\section{Acknowledgements}

We thank J. Seco for preliminary investigations of the discrepancy
which stimulated the re-analyses herein, A. Baptista for assistance
in last phases of the analysis, and G. Waysand for critical
comments. The work was supported in part by grants
PRAXIS/FIS/10033/1998 and POCTI/FNU/39067/2001 of the Foundation for
Science and Technology of Portugal, co-financed by FEDER.

%\bibliography{conmatprl}

\begin{thebibliography}{21}
\expandafter\ifx\csname
natexlab\endcsname\relax\def\natexlab#1{#1}\fi
\expandafter\ifx\csname bibnamefont\endcsname\relax
  \def\bibnamefont#1{#1}\fi
\expandafter\ifx\csname bibfnamefont\endcsname\relax
  \def\bibfnamefont#1{#1}\fi
\expandafter\ifx\csname citenamefont\endcsname\relax
  \def\citenamefont#1{#1}\fi
\expandafter\ifx\csname url\endcsname\relax
  \def\url#1{\texttt{#1}}\fi
\expandafter\ifx\csname urlprefix\endcsname\relax\def\urlprefix{URL
}\fi \providecommand{\bibinfo}[2]{#2}
\providecommand{\eprint}[2][]{\url{#2}}

\bibitem[{\citenamefont{Tinkham}(1975)}]{tinkh}
\bibinfo{author}{\bibfnamefont{M.}~\bibnamefont{Tinkham}},
  \emph{\bibinfo{title}{Introduction to Superconductivity}}
  (\bibinfo{publisher}{McGraw-Hill}, \bibinfo{address}{NY},
  \bibinfo{year}{1975}).

\bibitem[{\citenamefont{P.G. de Gennes}(1966)}]{degen}
\bibinfo{author}{\bibfnamefont{P.G.}~\bibnamefont{de Gennes}},
  \emph{\bibinfo{title}{Superconductivity of Metals and Alloys}}
  (\bibinfo{publisher}{W. A. Benjamin, Inc.}, \bibinfo{address}{NY},
  \bibinfo{year}{1966}).

\bibitem[{\citenamefont{Chang and Serin}(1966)}]{chang}
\bibinfo{author}{\bibfnamefont{G.~K.} \bibnamefont{Chang}} \bibnamefont{and}
  \bibinfo{author}{\bibfnamefont{B.}~\bibnamefont{Serin}},
  \bibinfo{journal}{Phys. Rev.} \textbf{\bibinfo{volume}{145}},
  \bibinfo{pages}{274} (\bibinfo{year}{1966}).

\bibitem[{\citenamefont{Miller and Cody}(1968)}]{miller}
\bibinfo{author}{\bibfnamefont{R.~E.} \bibnamefont{Miller}} \bibnamefont{and}
  \bibinfo{author}{\bibfnamefont{G.~D.} \bibnamefont{Cody}},
  \bibinfo{journal}{Phys. Rev.} \textbf{\bibinfo{volume}{173}},
  \bibinfo{pages}{494} (\bibinfo{year}{1968}).

\bibitem[{\citenamefont{Smith et~al.}(1970)}]{sbc}
\bibinfo{author}{\bibfnamefont{F.~W.}~\bibnamefont{Smith}},
  \bibinfo{author}{\bibfnamefont{A.} \bibnamefont{Baratoff}}, \bibnamefont{and}
  \bibinfo{author}{\bibfnamefont{M.} \bibnamefont{Cardona}},
  \bibinfo{journal}{Phys. Kondens. Materie}
  \textbf{\bibinfo{volume}{12}}, \bibinfo{pages}{145} (\bibinfo{year}{1970}).

\bibitem[{\citenamefont{de la Cruz et~al.}(1969)}]{cruz}
\bibinfo{author}{\bibfnamefont{F.}~\bibnamefont{de la Cruz}},
  \bibinfo{author}{\bibfnamefont{M.~D.} \bibnamefont{Maloney}}, \bibnamefont{and}
  \bibinfo{author}{\bibfnamefont{M.} \bibnamefont{Cardona}},
  \emph{\bibinfo{title}{Proc. Conf. on Science of Superconductivity}}
  (\bibinfo{publisher}{Stanford University Press}, \bibinfo{address}{Stanford},
  \bibinfo{year}{1969}).

\bibitem[{\citenamefont{de la Cruz et~al.}(1971)}]{cruz2}
\bibinfo{author}{\bibfnamefont{F.}~\bibnamefont{de la Cruz}},
  \bibinfo{author}{\bibfnamefont{M.~D.} \bibnamefont{Maloney}}, \bibnamefont{and}
  \bibinfo{author}{\bibfnamefont{M.} \bibnamefont{Cardona}},
  \bibinfo{journal}{Phys. Rev.}
  \textbf{\bibinfo{volume}{B3}}, \bibinfo{pages}{3802} (\bibinfo{year}{1971}).

\bibitem[{\citenamefont{Feder et~al.}(1966)}]{fedkiroth}
\bibinfo{author}{\bibfnamefont{J.}~\bibnamefont{Feder}},
  \bibinfo{author}{\bibfnamefont{S.~R.} \bibnamefont{Kiser}},
  \bibinfo{author}{\bibfnamefont{F.} \bibnamefont{Rothwarf}},
  \bibinfo{author}{\bibfnamefont{J.~P.} \bibnamefont{Burger}}, \bibnamefont{and}
  \bibinfo{author}{\bibfnamefont{C.} \bibnamefont{Valette}},
  \bibinfo{journal}{Sol. State. Comm.}
  \textbf{\bibinfo{volume}{4}}, \bibinfo{pages}{611} (\bibinfo{year}{1966}).

\bibitem[{\citenamefont{Smith et~al}(1968)}]{sc}
\bibinfo{author}{\bibfnamefont{F.~W.}~\bibnamefont{Smith}} \bibnamefont{and}
  \bibinfo{author}{\bibfnamefont{M.}~\bibnamefont{Cardona}},
  \bibinfo{journal}{Sol. State. Comm.} \textbf{\bibinfo{volume}{6}},
  \bibinfo{pages}{37} (\bibinfo{year}{1968}).

\bibitem[{\citenamefont{Burger et~al.}(1994)}]{burger}
\bibinfo{author}{\bibfnamefont{J.~P.}~\bibnamefont{Burger}},
  \bibinfo{author}{\bibfnamefont{J.} \bibnamefont{Feder}}, \bibnamefont{and}
  \bibinfo{author}{\bibfnamefont{S.~R.} \bibnamefont{Kiser}},
  \bibnamefont{et~al.}, \emph{\bibinfo{title}{Proc. 10th Int'l Conf. on Low Temperature Physics}}
  (\bibinfo{publisher}{VINITI}, \bibinfo{address}{Moscow},
  \bibinfo{pages}{352} \bibinfo{year}{1967}).

\bibitem[{\citenamefont{Feder and McLachlan}(1968)}]{feder}
\bibinfo{author}{\bibfnamefont{J.}~\bibnamefont{Feder}} \bibnamefont{and}
  \bibinfo{author}{\bibfnamefont{D.}~\bibnamefont{McLachlan}},
  \bibinfo{journal}{Phys. Rev.} \textbf{\bibinfo{volume}{177}},
  \bibinfo{pages}{763} (\bibinfo{year}{1968}).

\bibitem[{\citenamefont{Feder et~al.}(1994)}]{fedkiroth2}
\bibinfo{author}{\bibfnamefont{J.} \bibnamefont{Feder}},
  \bibinfo{author}{\bibfnamefont{S.~R.} \bibnamefont{Kiser}}, \bibnamefont{and}
  \bibinfo{author}{\bibfnamefont{F.} \bibnamefont{Rothwarf}},
  \bibinfo{journal}{Phys. Rev. Lett.} \textbf{\bibinfo{volume}{17}},
  \bibinfo{pages}{87} (\bibinfo{year}{1966}).

\bibitem[{\citenamefont{Smith et~al}(1968)}]{sc2}
\bibinfo{author}{\bibfnamefont{F.~W.}~\bibnamefont{Smith}} \bibnamefont{and}
  \bibinfo{author}{\bibfnamefont{M.}~\bibnamefont{Cardona}},
  \bibinfo{journal}{Sol. State. Comm.} \textbf{\bibinfo{volume}{5}},
  \bibinfo{pages}{345} (\bibinfo{year}{1966}).

\bibitem[{\citenamefont{Larrea et~al}(1968)}]{larrea}
\bibinfo{author}{\bibfnamefont{A.}~\bibnamefont{Larrea}}
  \bibnamefont{et~al.},
  \bibinfo{journal}{Nucl. Instr. \& Meth.} \textbf{\bibinfo{volume}{A137}},
  \bibinfo{pages}{541} (\bibinfo{year}{1992}).

\bibitem[{\citenamefont{Smith et~al}(1967)}]{sc3}
\bibinfo{author}{\bibfnamefont{F.~W.}~\bibnamefont{Smith}} \bibnamefont{and}
  \bibinfo{author}{\bibfnamefont{M.}~\bibnamefont{Cardona}},
  \bibinfo{journal}{Phys. Lett.} \textbf{\bibinfo{volume}{24A}},
  \bibinfo{pages}{247} (\bibinfo{year}{1967}).

\bibitem[{\citenamefont{Smith et~al}(1967)}]{sc4}
\bibinfo{author}{\bibfnamefont{F.~W.}~\bibnamefont{Smith}} \bibnamefont{and}
  \bibinfo{author}{\bibfnamefont{M.}~\bibnamefont{Cardona}},
  \bibinfo{journal}{Phys. Lett.} \textbf{\bibinfo{volume}{25A}},
  \bibinfo{pages}{345} (\bibinfo{year}{1967}).

\bibitem[{\citenamefont{Valko et~al.}(1994)}]{valko}
\bibinfo{author}{\bibfnamefont{P.}~\bibnamefont{Valko}},
  \bibinfo{author}{\bibfnamefont{M.~R.} \bibnamefont{Gomes}}, \bibnamefont{and}
  \bibinfo{author}{\bibfnamefont{TA} \bibnamefont{Girard}},
  \bibinfo{journal}{Phys. Rev.}
  \textbf{\bibinfo{volume}{B75}}, \bibinfo{pages}{140504(R)} (\bibinfo{year}{2007}).

\bibitem[{\citenamefont{Auer and Ullmaier}(1973)}]{auer}
\bibinfo{author}{\bibfnamefont{J.}~\bibnamefont{Auer}} \bibnamefont{and}
  \bibinfo{author}{\bibfnamefont{H.}~\bibnamefont{Ullmaier}},
  \bibinfo{journal}{Phys. Rev.} \textbf{\bibinfo{volume}{B66}},
  \bibinfo{pages}{136} (\bibinfo{year}{1973}).

\bibitem[{\citenamefont{Landau and Ott}(2003)}]{lan}
\bibinfo{author}{\bibfnamefont{I.~L.} \bibnamefont{Landau}}, \bibnamefont{and}
  \bibinfo{author}{\bibfnamefont{H.~R.} \bibnamefont{Ott}},
  \bibinfo{journal}{Physica} \textbf{\bibinfo{volume}{C398}},
  \bibinfo{pages}{73} (\bibinfo{year}{2003}).

\bibitem[{\citenamefont{Ebneth and Tewordt}(1965)}]{ebneth}
\bibinfo{author}{\bibfnamefont{G.}~\bibnamefont{Ebneth}}, \bibnamefont{and}
  \bibinfo{author}{\bibfnamefont{L.} \bibnamefont{Tewordt}},
  \bibinfo{journal}{Z. Physik} \textbf{\bibinfo{volume}{185}},
  \bibinfo{pages}{421} (\bibinfo{year}{1965}).

\bibitem[{\citenamefont{Ginzburg}(1958)}]{ginz}
\bibinfo{author}{\bibfnamefont{V.~L.}~\bibnamefont{Ginzburg}},
  \bibinfo{journal}{Sov. Phys. JETP} \textbf{\bibinfo{volume}{7}},
  \bibinfo{pages}{78} (\bibinfo{year}{1958}).

\bibitem[{\citenamefont{Thompson and Baratoff}(1968)}]{tombara}
\bibinfo{author}{\bibfnamefont{R.~S.}~\bibnamefont{Thompson}}, \bibnamefont{and}
  \bibinfo{author}{\bibfnamefont{A.} \bibnamefont{Baratoff}},
  \bibinfo{journal}{Phys. Rev.} \textbf{\bibinfo{volume}{167}},
  \bibinfo{pages}{361} (\bibinfo{year}{1968}).

\bibitem[{\citenamefont{Dolgert, Bartolo and Dorsey}(1996)}]{dolgert}
\bibinfo{author}{\bibfnamefont{A.~J.}~\bibnamefont{Dolgert}},
  \bibinfo{author}{\bibfnamefont{S.~J.~D.} \bibnamefont{Bartolo}}, \bibnamefont{and}
  \bibinfo{author}{\bibfnamefont{A.~T.} \bibnamefont{Dorsey}},
  \bibinfo{journal}{Phys. Rev.} \textbf{\bibinfo{volume}{B534}},
  \bibinfo{pages}{5650} (\bibinfo{year}{1996-I}).

\bibitem[{\citenamefont{Hueber,Valette and Waysand}(1979)}]{hueber}
\bibinfo{author}{\bibfnamefont{D.}~\bibnamefont{Hueber}},
  \bibinfo{author}{\bibfnamefont{C.} \bibnamefont{Valette}}, \bibnamefont{and}
  \bibinfo{author}{\bibfnamefont{G.} \bibnamefont{Waysand}},
  \bibinfo{journal}{Physica} \textbf{\bibinfo{volume}{108B}},
  \bibinfo{pages}{1229} (\bibinfo{year}{1979}).

\bibitem[{\citenamefont{Penaranda, Auguet and Ramirez-Piscina}(1999)}]{pen1}
\bibinfo{author}{\bibfnamefont{A.} \bibnamefont{Penaranda}},
  \bibinfo{author}{\bibfnamefont{C.~E.} \bibnamefont{Auguet}}, \bibnamefont{and}
  \bibinfo{author}{\bibfnamefont{L.} \bibnamefont{Ramirez-Piscina}},
  \bibinfo{journal}{Nucl. Instr. \& Meth.} \textbf{\bibinfo{volume}{A424}},
  \bibinfo{pages}{512} (\bibinfo{year}{1999}).

\bibitem[{\citenamefont{Caddah and Roy}(1999)}]{chaddah}
\bibinfo{author}{\bibfnamefont{P.} \bibnamefont{Chaddah}}, \bibnamefont{and}
  \bibinfo{author}{\bibfnamefont{S.~B.} \bibnamefont{Roy}},
  \bibinfo{journal}{Phys. Rev.} \textbf{\bibinfo{volume}{B60}},
  \bibinfo{pages}{11926} (\bibinfo{year}{1999-I}).

\bibitem[{\citenamefont{Hove and Sudb{\o}}(2002)}]{hove}
\bibinfo{author}{\bibfnamefont{S.~M.} \bibnamefont{J.~Hove}} \bibnamefont{and}
  \bibinfo{author}{\bibfnamefont{A.}~\bibnamefont{Sudb{\o}}},
  \bibinfo{journal}{Phys. Rev.} \textbf{\bibinfo{volume}{B66}},
  \bibinfo{pages}{064524} (\bibinfo{year}{2002}).

\end{thebibliography}

\end{document}